\documentclass[a4paper,11pt]{article}
\usepackage{jinstpub} 
\usepackage{lineno}

\usepackage[separate-uncertainty = true]{siunitx}
\usepackage{subcaption}
\usepackage{placeins}
\usepackage{cleveref}
\usepackage{multicol}

\proceeding{25$^{\text{th}}$ International Workshop on Radiation Imaging Detectors\\
Lisbon, Portugal\\
30 June - 4 July 2024}

\title{\boldmath{Laboratory and beam-test performance study of a 55~\textmu m pitch iLGAD sensor bonded to a Timepix3 readout chip}}

\author[a,b,1]{P. Svihra,\note{Corresponding author.}}
\author[c]{R. Bates}
\author[a]{J. Braach}
\author[a,d]{E. Buschmann}
\author[a]{D. Dannheim}
\author[c]{D. Maneuski}
\author[e]{N. Moffat}
\author[a]{Y. Otarid}

\affiliation[a]{EP-DT-TP, CERN, Esplanade des Particules, Meyrin, Switzerland}
\affiliation[b]{Department of Detector Development and Data Processing, Institute of Physics of the Czech Academy of Sciences, Na Slovance, Prague, Czech Republic}
\affiliation[c]{School of Physics and Astronomy, University of Glasgow, Glasgow, United Kingdom}
\affiliation[d]{Physics Department, BNL, Upton, NY, United States of America}
\affiliation[e]{Radiation Detectors Group, IMB-CNM-CSIC, Carrer dels Til·lers, Barcelona, Spain}

\emailAdd{peter.svihra@cern.ch}

\abstract{This contribution reports on characterisation results of a large-area (2~cm\textsuperscript{2}) small pitch (55~\textmu m) inverse Low-Gain Avalanche Detector (iLGAD), bonded to a Timepix3 readout chip. The ilGAD sensors were produced by Micron Semiconductor Ltd with the goal to obtain good gain uniformity and maximise the fill-factor -- an issue present with standard small-pitch LGAD designs. We have conducted detailed performance evaluations using both X-ray calibrations and beam tests. An X-ray fluorescence setup has been used to obtain energy calibration and to identify the optimal operating settings of the new devices, whereas the extensive beam tests allowed for a detailed evaluation of the detector performance. The beam-tests were performed at the CERN SPS North Area H6 beamline, using a 120 GeV/c pion beam. The reference tracking and time-stamping is achieved by a Timepix3-based beam telescope setup.

The results show a gain of around 5 with very good uniformity, measured across the whole gain area, as well as a hit time resolution down to 1.3 ns without correcting for the time-walk effects. Furthermore, it is shown that the gain opens the possibility of a good X-ray energy resolution down to 4.5 keV.}


\arxivnumber{2409.20194} 

\begin{document}
\maketitle
\flushbottom

\section{Introduction and motivation}
Low Gain Avalanche Detectors (LGAD) are recent types of semiconductor detectors that improve the signal formation in sensors using multiplication of electron-hole pairs by a factor around 10 \cite{lgad}.
The gain is achieved by implanting high concentrations of p- and n- type dopants in the silicon bulk, resulting in strong electric fields $\mathcal O(\SI{1e15}{\volt\per\centi\meter})$ within the sensor that triggers impact ionisation of free charge carriers \cite{ionisation}.
Such an approach enables the sensors to be thin (about \SI{50}{\micro\meter)}, lowering the material budget and at the same time decreasing the distance required for the charge collection thus improving the timing for minimum ionising particles (MIPs) down to below \SI{50}{\pico\second} \cite{timing}.
The High-Luminosity upgrades of the ATLAS and CMS experiments at CERN will contain LGAD-based MIP timing layers with $\mathcal O(\SI{1}{\milli\meter\squared})$ pixel sizes for improving the separation of tracks in time \cite{atlas,cms} but charge multiplication can enable the detection of low-energy X-rays below the readout threshold \cite{xray}.

\begin{figure}[t]
    \begin{subfigure}[t]{0.48\textwidth}
        \centering
        \includegraphics[width=1.\linewidth]{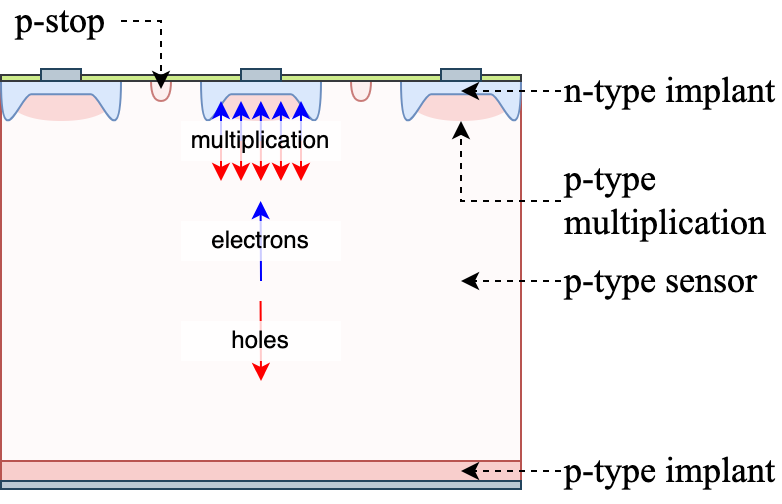}
        \caption{LGAD -- multiplication in each pixel, need of field separation using JTE.}
        \label{fig:lgad}
    \end{subfigure}
    \hfill
    \begin{subfigure}[t]{0.48\textwidth}
        \centering
        \includegraphics[width=1.\linewidth]{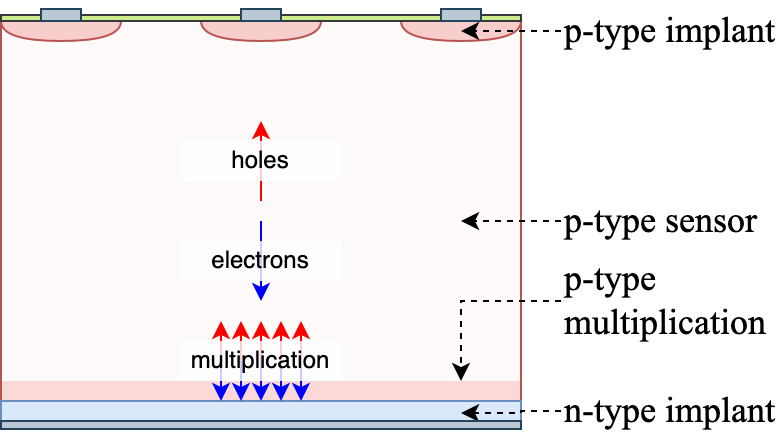}
        \caption{iLGAD -- flat multiplication layer at the back-side.}
        \label{fig:ilgad}
    \end{subfigure}
    \caption{Visual comparison of the standard and inverse LGAD design.}
\end{figure}

\begin{figure}[t]
    \centering
    \includegraphics[width=0.65\linewidth]{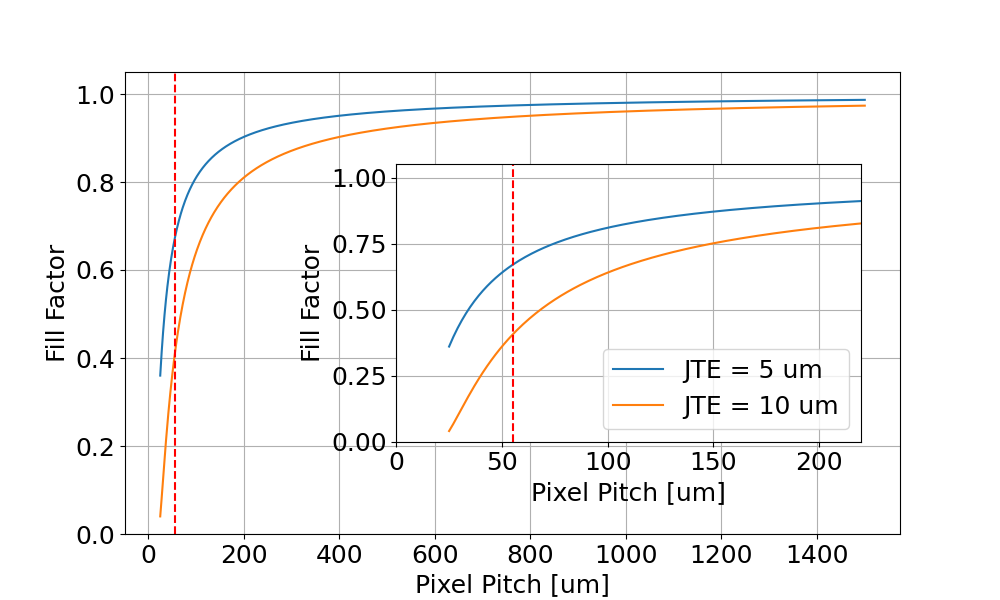}
    \caption{Comparison of geometrical fill factor $F$ for different widths of junction terminations (JTE). The red dotted line indicates the \SI{55}{\micro\meter} pixel pitch matching the Timepix3 readout.}
    \label{fig:fill_factor}
\end{figure}

For many high-energy physics applications the most commonly used type of silicon sensor bulk is p-type which is targets electron collection in the readout pixels.
These sensors are often produced with multiplication layers directly underneath each pixel, with the added requirement of proper termination of the strong electric fields using either Junction Termination (JTE), shown in \cref{fig:lgad}, or trenches filled with oxide.
A downside of this approach is seen when decreasing the pixel pitch, where the no-gain pixel area starts to affect the signal formation.
This effect can be approximated by a geometrical fill factor $F$ described as
\begin{equation}
    F = \frac{A_\mathrm{gain}}{A_\mathrm{total}},
\end{equation}
where $A$ represents gain and total areas covered by a pixel.
The dependency on the size of the no-gain region caused by the JTE is known \cite{fill_factor}, demonstrated in \cref{fig:fill_factor}.
There are different approaches that aim to mitigate this effect - this work focuses on the characterisation of iLGAD sensors \cite{ilgad}.

\subsection{Sensor design}
The sensor has been produced by Micron Semiconductor Ltd as part of R\&D efforts by Universities of Manchester, Glasgow and Edinburgh.
Compared to the standard LGAD design, an iLGAD sensor uses the same p-type bulk, but the multiplication layer is at the backside (schematic shown in \cref{fig:ilgad}).
This design results in a uniform multiplication across the full sensor at the cost of double-sided processing, hole collection at the pixels and need for full depletion.

The produced sensors are \SI{300}{\micro\meter} thick with 256x256 pixels with \SI{55}{\micro\meter} pitch and have been bump-bonded to Timepix3 readout chips.
In order to simplify the gain and no-gain performance evaluation, an outer rim 4 pixels wide has been included without multiplication layer, followed by a pixel-wide rim containing JTE and finally an area of 246x246 pixels with multiplication.

\subsection{Timepix3 readout chip}
The Timepix3 versatile readout chip \cite{timepix3} has a pixel matrix of 256x256 pixels with \SI{55}{\micro\meter} pitch.
The zero-suppressed data-driven readout provides access to a Time over Threshold (ToT) and Time of Arrival (ToA) information.
While ToT provides a measurement of the deposited charge, ToA measures the time of the signal crossing the threshold with a binning of \SI{1.5625}{\nano\second} -- making the chip suitable for evaluating the charge collection of the new sensor designs and for performing an initial characterisation of the timing performance.

The chip is read out using the SPIDR DAQ system \cite{SPIDR}.
Apart from the raw Timepix3 data, the readout also provides a Time-to-digital converter (TDC) input with binning of \SI{260}{\pico\second}.

\section{Laboratory characterisation}
As part of evaluating the performance of the device, a set of laboratory tests has been performed.
The testing presented here was done after bump-bonding the sensor to an ASIC.
Both parts were confirmed to be of good quality based on wafer-level examinations.
Results from a single iLGAD sensor bonded to a Timepix3 chip (ID W0068\_I11) are presented.

The tests included a current vs voltage scan to establish the operating points of the device, approximating depletion voltage of the gain layer $V_\mathrm{gain} = \SI{25}{\volt}$ and about $V_\mathrm{dep} = \SI{150}{\volt}$ for the full sensor.
The threshold response of the pixels in the Timepix3 chip has been then equalised based on the electronic noise, resulting in 114 masked pixels (\SI{0.2}{\percent}) due to increased noise.
The average noise baseline was estimated to \SI{1088.5(2.4)}{DAC} units with an average electronic noise of \SI{11.5(1.4)}{DAC} units.
The default operational threshold has been set to noise-free operation at \SI{1000}{DAC} and was used for all main measurements.
Furthermore the device has been calibrated using the on-chip available test-pulse injection and finally an energy calibration using X-rays.

\subsection{Test-pulse energy calibration}
The Timepix3 chip also has the capability of injecting analogue and digital pulses to each pixel.
Using this approach, the signal amplification, shaping and digitisation can be tested without the need of radioactive sources, resulting in measurements of ToT and ToA at arbitrary input energies.

The injected energy $E' = CV$ is obtained by charging the built-in capacitor $C\approx\SI{3.2}{\femto\farad}$ to different voltages $V$.
The value of the test-pulse capacitor is constrained by design, but cannot be measured directly.
An uncertainty of $\pm\SI{20}{\percent}$ is assumed caused by variations in the semiconductor manufacturing processes.
By discharging the capacitor and measuring the ToT response in every pixel, a dependency in the form
\begin{equation}
\label{eq:calibration_testpulse}
ToT = A'E' + b' - \frac{c'}{E' - t'}
\end{equation}
is observed, where $A'$ and $b'$ correspond to the gain and offset of the linear part and $c'$ and $t'$ describe the curvature and asymptote of the non-linear part observed at small amplitudes.

\subsection{X-ray energy calibration}
Following the initial test-pulse energy calibration, X-rays of known energy are used for two purposes: providing conversion from threshold DAC units to energy as well as obtaining the per-pixel ToT energy response.
Both tests were done utilising an X-ray tube with a copper target and the resulting X-rays hitting secondary (changeable) targets producing X-ray fluorescence (XRF) peaks.
Materials used for the targets were titanium (\SI{4.51}{\kilo\electronvolt}), iron (\SI{6.40}{\kilo\electronvolt}), copper (\SI{8.05}{\kilo\electronvolt}), molybdenum (\SI{17.48}{\kilo\electronvolt}), silver (\SI{22.16}{\kilo\electronvolt}) and tin (\SI{25.27}{\kilo\electronvolt}).
The quoted energies represent the respective K$\alpha$1 edges.
To minimise any charge-sharing effects, only single pixel clusters were analysed.

\begin{figure}[t]
    \centering
    \includegraphics[width=0.75\linewidth]{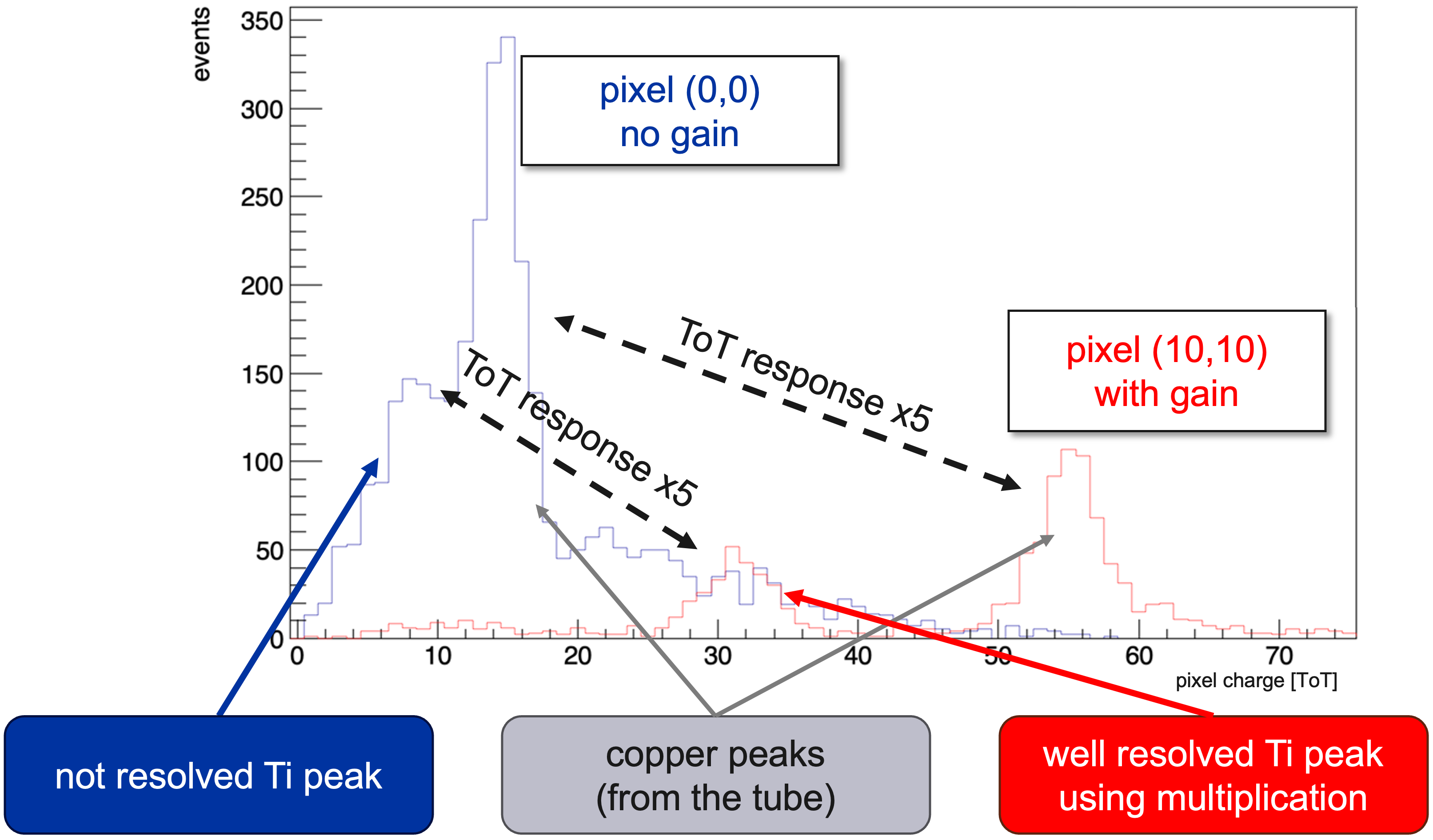}
    \caption{Example of the detector single-pixel XRF response for pixels without and with gain. Gain of around 5 as well as improved low-energy resolution is clearly seen.}
    \label{fig:example_spectrum}
\end{figure}

\begin{figure}[t]
    \begin{subfigure}[t]{0.49\textwidth}
        \centering
        \includegraphics[width=1.\linewidth]{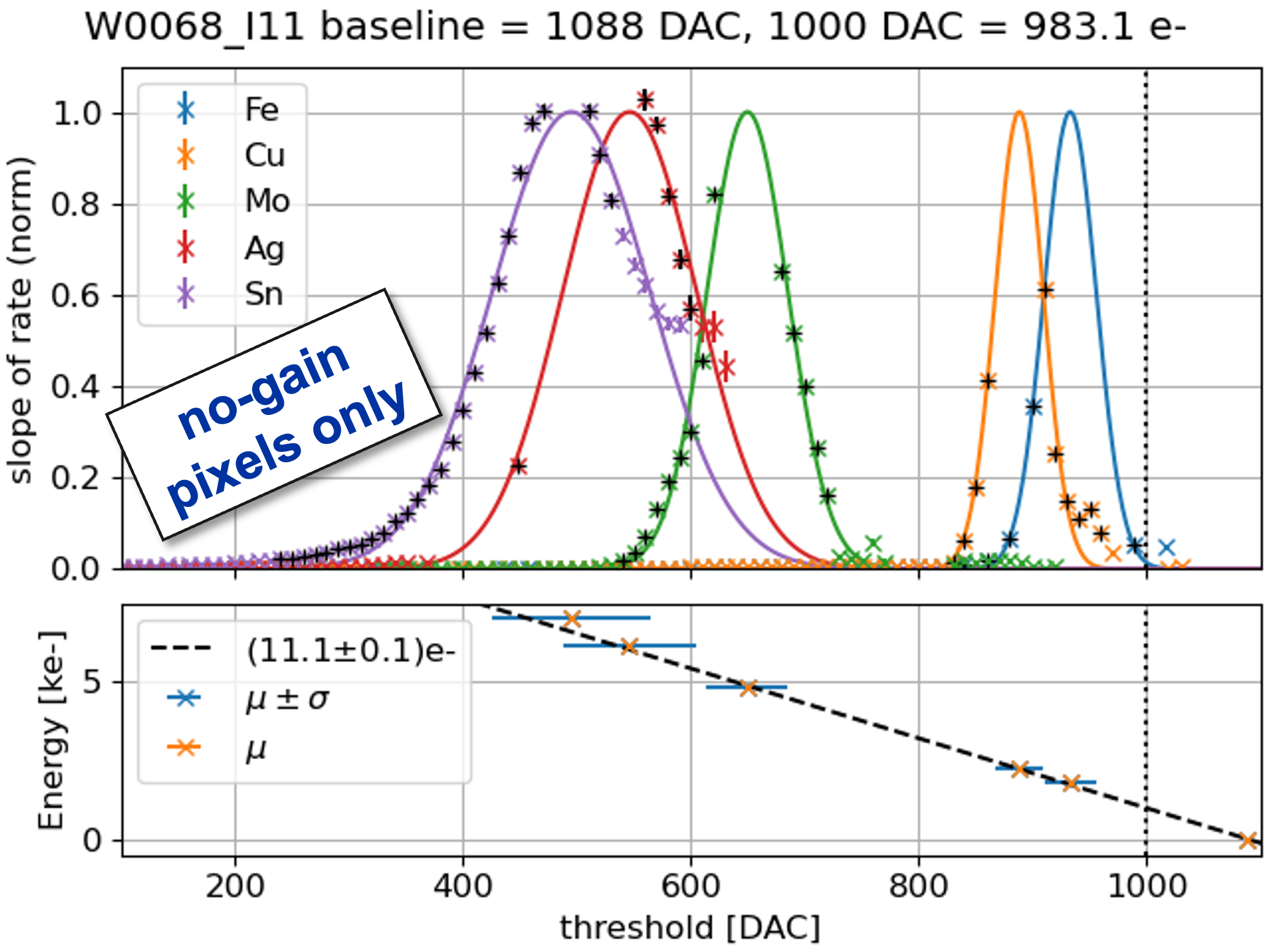}
        \caption{Threshold calibration.}
        \label{fig:calibration_threshold}
    \end{subfigure}
    \hfill
    \begin{subfigure}[t]{0.49\textwidth}
        \centering
        \includegraphics[width=1.\linewidth]{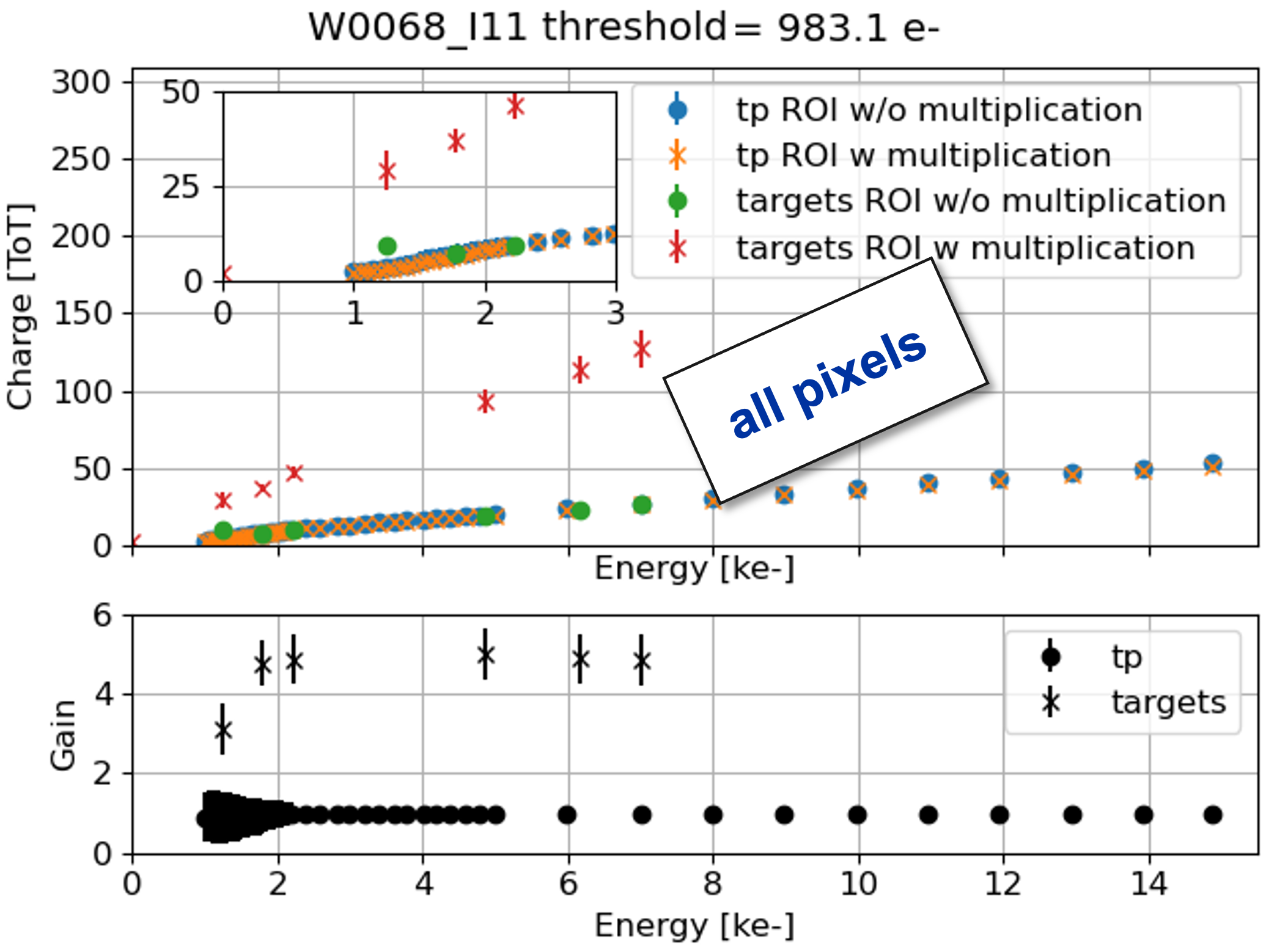}
        \caption{Pixel energy calibration.}
        \label{fig:calibration_energy}
    \end{subfigure}
    \caption{Charge calibration results done at \SI{200}{\volt}.}
\end{figure}

In order to evaluate the global threshold response, only the outer rim of pixels bonded to the no-gain parts of the sensor was used for the analysis.
By fixing the bias voltage to $V = \SI{200}{\volt}$ and taking data at different threshold values for each individual XRF target, hit rate curves were obtained.
The derivatives of the measurements are shown in \cref{fig:calibration_threshold}.
Each peak corresponds to a specific energy and is fitted with a gaussian.
The threshold response is obtained by a linear fit of the means, resulting in a conversion of \SI{11.1(0.1)}{e-} per \SI{1}{DAC} unit.

Afterwards, by fixing the operational threshold to \SI{983}{e-} (corresponding to \SI{1000}{DAC}) and acquiring more data, a larger sample has been obtained enabling the measurement of the ToT pixel response (example result shown in \cref{fig:example_spectrum}).
\Cref{fig:calibration_energy} shows a comparison of the averaged ToT response for pixels from both the gain and no-gain regions, as well as the corresponding testpulse response.
The plot also shows that the \SI{4.86}{\kilo\electronvolt} energy peak from titanium is measurable only by pixels with gain, whereas the signal in no-gain pixels is dominated by the higher energy copper peak caused by X-rays originating in the tube.
This is better seen by plotting the ratio of the averaged ToT values where for all peak energies, apart from titanium, a uniform gain of 
\begin{equation}
    G = (4.86\pm0.08),
\end{equation}
is obtained.
In comparison, the ratios of the same ToT averaged pixel areas for the testpulse injection remain at 1.
This confirms that the multiplication layer in part of the sensor does not impact the readout electronics and only increases the measured sensor signal.

Based on the results, a final per-pixel energy correction was obtained in the form
\begin{equation}
    ToT = AE + b - \frac{\frac{c'A'}{A}}{E - \frac{A't' +b' - b}{A}},
\end{equation}
where $A$ and $b$ are linear components of the dependency of $ToT$ on X-ray energy $E$ and the other parameters come from the testpulse calibration \cref{eq:calibration_testpulse}.
The calibration in this form has been applied to all subsequent beam-test measurements.

\section{Beam-test characterisation}
To characterise the MIP response of the assembly a beam-test campaign has been performed.
For the tests a pion beam with momentum \SI{120}{\giga\electronvolt\per c} at the H6 beamline at the CERN SPS North Area has been used.
The tracks were measured by the CLICdp Timepix3 beam telescope \cite{telescope}.

Tracking is achieved by utilising 3 upstream and 3 downstream Timepix3 detector planes tilted by 9 degrees along two axes and upstream and downstream scintillators with photomultiplier tubes for track timestamping.
The pointing resolution at the centre of the telescope in the device under test (DUT) location is about \SI{1.5}{\micro\meter} and the track timing resolution is about \SI{300}{\pico\second}.
Reconstruction and analysis of the data was performed using the Corryvreckan framework \cite{corry}.

\subsection{Charge collection and efficiency}
Since the iLGAD DUT is larger than the acceptance of the beam telescope, all measurements were repeated in 4 positions such that every corner of the DUT was placed in the centre of the beam.

\begin{figure}[t]
    \begin{minipage}[t]{0.48\linewidth}
        \centering
        \includegraphics[width=1.\linewidth]{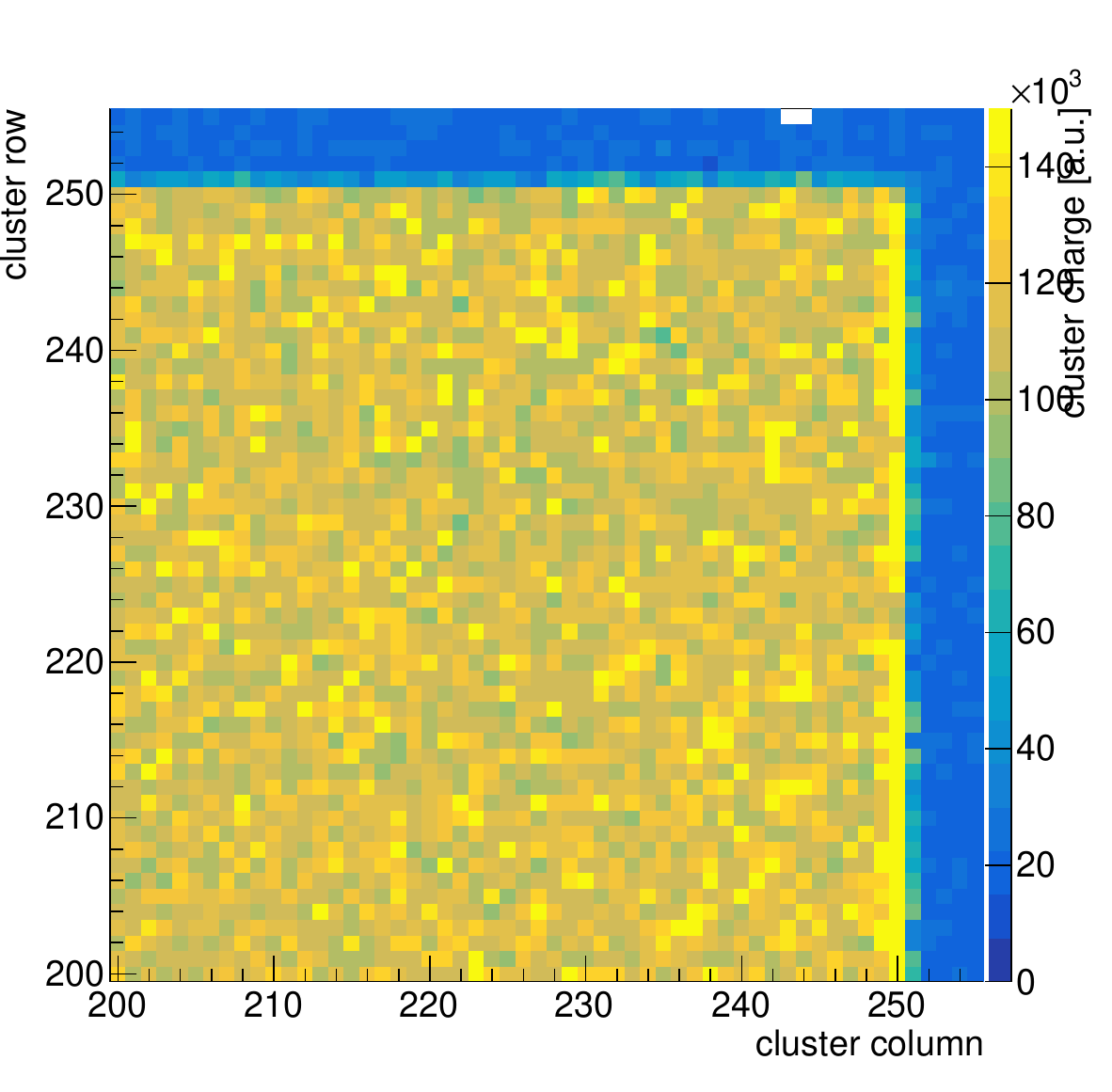}
        \caption{Measured cluster charges for bias \SI{300}{\volt} and threshold \SI{1}{\kilo e-} in one corner.}
        \label{fig:chip-charge}        
    \end{minipage}
    \hfill
    \begin{minipage}[t]{0.48\linewidth}
        \centering
        \includegraphics[width=1.\linewidth]{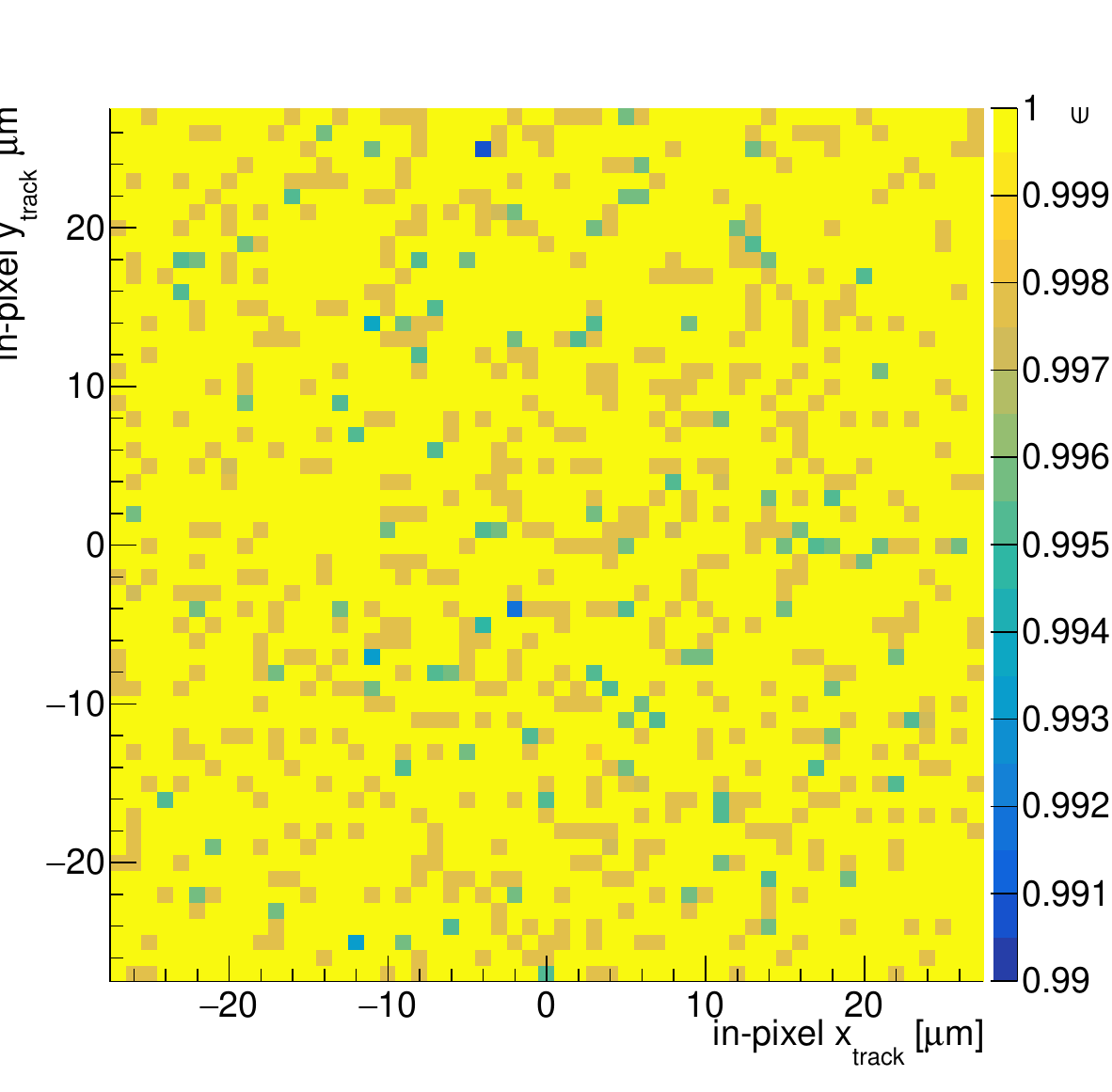}
        \caption{In-pixel efficiencies mapped to single pixel for bias \SI{300}{\volt} and threshold \SI{1}{\kilo e-}.}
        \label{fig:chip-eff}
    \end{minipage}
\end{figure}

Firstly, we have evaluated the signal response and in-pixel efficiency of signal detection at the operational point of \SI{300}{\volt} and \SI{983}{e-}, as shown in \cref{fig:chip-charge} and \cref{fig:chip-eff}, respectively.
Excellent multiplication and detection performance is visible in both plots, overall efficiency is above \SI{99.94}{\percent}.
The outer rim in cluster charge matches with the region with pixels without multiplication layer.

\begin{figure}[t]
    \begin{subfigure}[t]{0.45\textwidth}
        \centering
        \includegraphics[width=1.\linewidth]{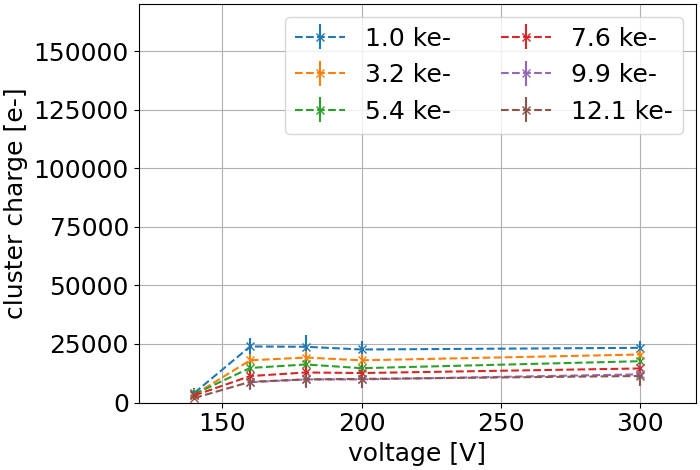}
        \caption{No-gain.}
    \end{subfigure}
    \hfill
    \begin{subfigure}[t]{0.44\textwidth}
        \centering
        \includegraphics[width=1.\linewidth]{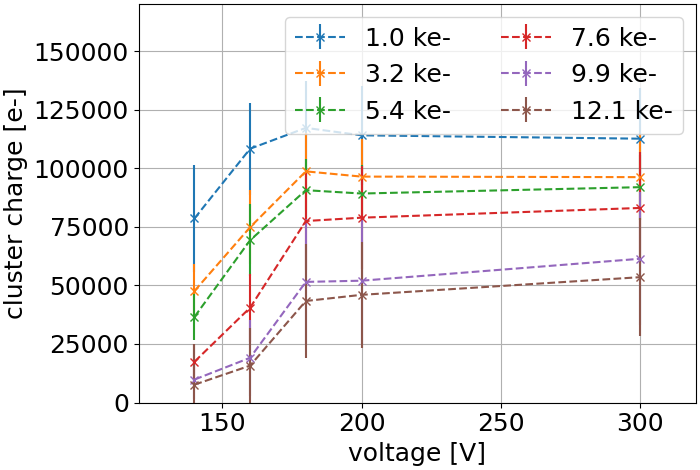}
        \caption{Gain.}
    \end{subfigure}
    \caption{Measured average cluster charge as a function of bias voltage and threshold, compared for pixels without and with multiplication. Error bars are RMS of the distributions.}
    \label{fig:beam-charge}
\end{figure}

\begin{figure}[t]
    \begin{subfigure}[t]{0.45\textwidth}
        \centering
        \includegraphics[width=1.\linewidth]{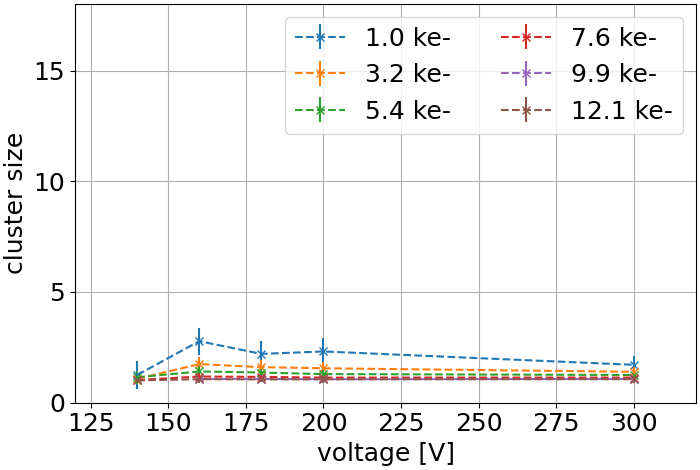}
        \caption{No-gain.}
    \end{subfigure}
    \hfill
    \begin{subfigure}[t]{0.45\textwidth}
        \centering
        \includegraphics[width=1.\linewidth]{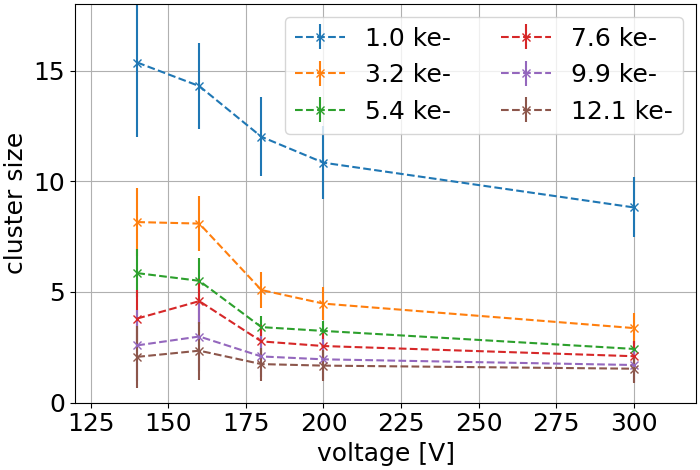}
        \caption{Gain}
    \end{subfigure}
    \caption{Measured average cluster size as a function of bias voltage and threshold, compared for pixels without and with multiplication. Error bars are RMS of the distributions.}
    \label{fig:beam-size}
\end{figure}

\begin{figure}[t]
    \begin{subfigure}[t]{0.45\textwidth}
        \centering
        \includegraphics[width=1.\linewidth]{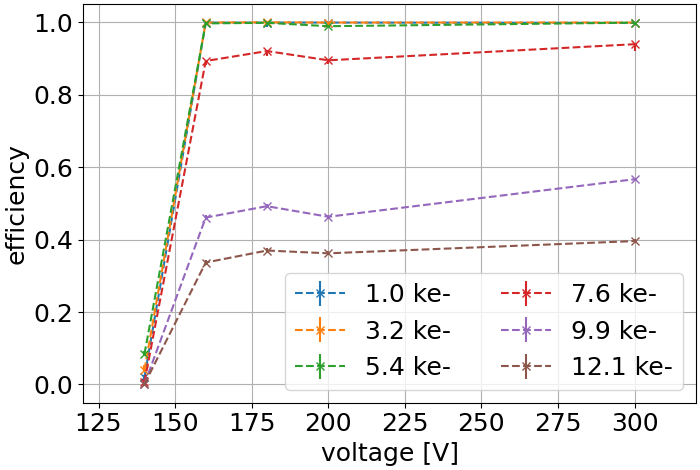}
        \caption{No-gain.}
    \end{subfigure}
    \hfill
    \begin{subfigure}[t]{0.45\textwidth}
        \centering
        \includegraphics[width=1.\linewidth]{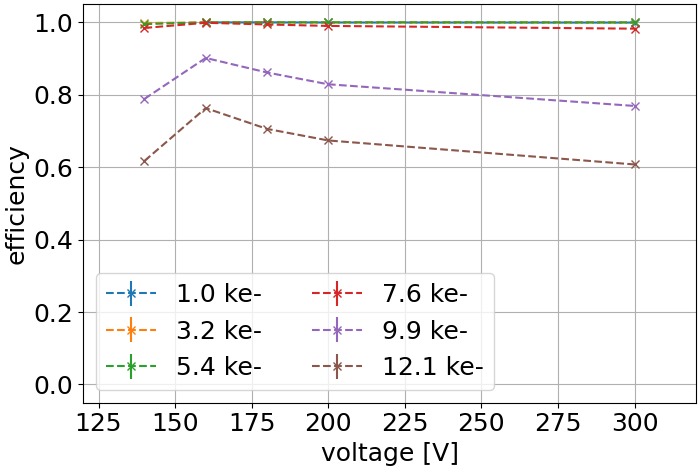}
        \caption{Gain.}
    \end{subfigure}
    \caption{Measured average tracking efficiency as a function of bias voltage and threshold, compared for pixels without and with multiplication. Efficiency for voltages below \SI{150}{\volt} drops off as the device is not fully depleted and the depletion starts at backside.}
    \label{fig:beam-eff}
\end{figure}

Then we have performed a combined voltage-threshold scan, covering the operational voltages from \SIrange{140}{300}{\volt} and thresholds from \SIrange{1}{12}{\kilo e-}.
Comparison plots of pixels without and with multiplication are in \cref{fig:beam-charge,fig:beam-size,fig:beam-eff}.
An impact of the gain can be seen, increasing the efficiency, charge and cluster size.
The latter is caused by diffusion of a large number of holes along their drift path in the \SI{300}{\micro\meter} thick silicon, mainly created in the multiplication layer on the backside.

\subsection{Gain and timing}
\begin{figure}[t]
    \begin{minipage}[t]{0.45\linewidth}
        \centering
        \includegraphics[width=1.\linewidth]{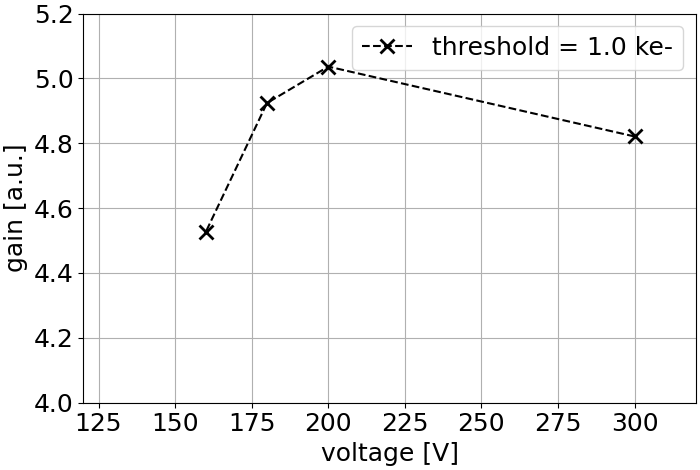}
        \caption{Calculated gain as a ratio of cluster charge measurements}
        \label{fig:gain}
    \end{minipage}
    \hfill
    \begin{minipage}[t]{0.45\linewidth}
        \includegraphics[width=1.\linewidth]{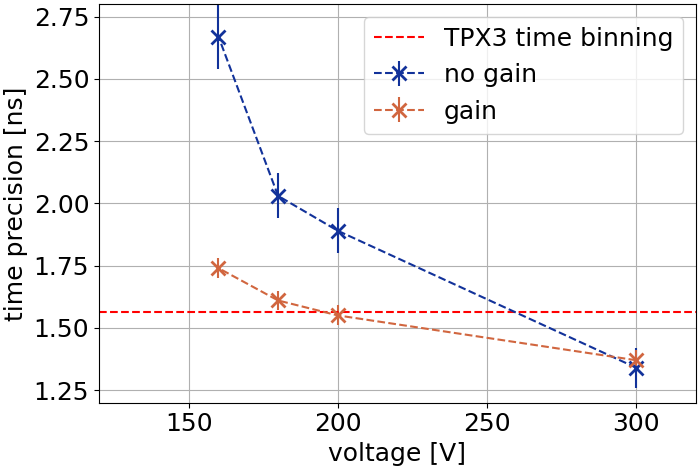}
        \caption{Time precision for threshold \SI{1}{\kilo e-}, without time-walk correction.}
        \label{fig:timing}
    \end{minipage}
\end{figure}

Similarly to the X-ray calibration, the ratio of the average cluster charge from pixels with and without multiplication in the sensor has been calculated.
Such a gain estimate has been done as a function of bias voltage for the nominal operational threshold, results as shown in \cref{fig:gain}.
An increase of gain is observed for voltages in the range from \SIrange{160}{200}{\volt} after which the gain decreases.
The decrease could be explained by a non-linear calibration result of the Timepix3 chip for very large signals, beyond the reach of the performed calibration \cite{volcano}, or by onset of a charge screening \cite{screening}.

To evaluate the timing performance, the residuals between the cluster times and the track times provided by the telescope were fitted with a Gaussian.
The resulting voltage dependency of the Gaussian width values, after quadratic subtraction of the telescope track-time resolution, is shown in \cref{fig:timing}.
The performance of the pixels with gain is overall better than without gain.
The observed timing performance is expected to be limited by the sensor design (hole-collection), the front-end electronics, and time-walk effects that were not corrected for.

\subsection{Angular scan}
Finally, the dependency of the track reconstruction on the incidence angle has been measured, providing information on potential charge screening for perpendicular incidence angles.
The used telescope setup enables a precise rotation of the DUT along the vertical axis, the angle has been recorded as angle between the beam and the nominal of the sensor -- a perpendicular track is therefore denoted as \ang{0}.

\begin{figure}[t]
    \begin{subfigure}[t]{0.45\linewidth}
        \centering
        \includegraphics[width=1.\linewidth]{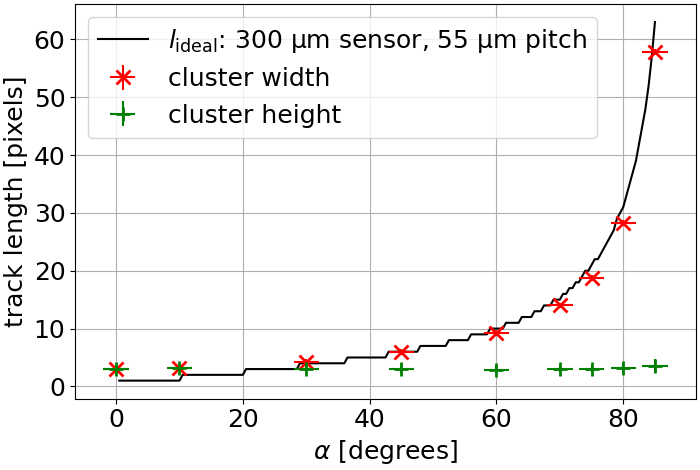}
        \caption{Cluster size.}
        \label{fig:angular_size}    
    \end{subfigure}
    \hfill
    \begin{subfigure}[t]{0.45\linewidth}
        \centering
        \includegraphics[width=1.\linewidth]{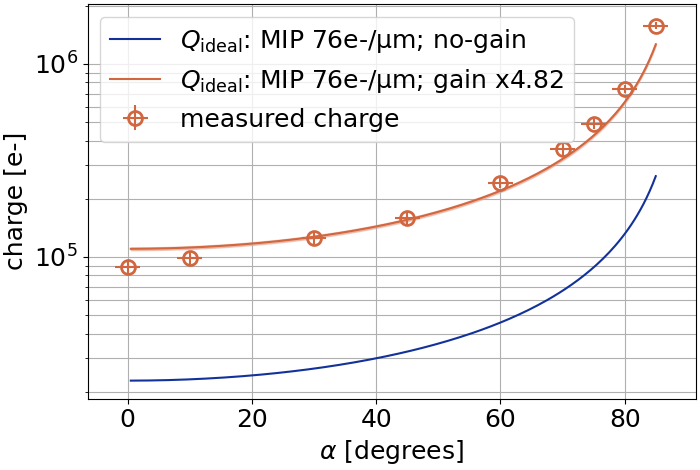}
        \caption{Charge.}
        \label{fig:angular_charge}
    \end{subfigure}
    \caption{Comparison of measured and calculated track length and collected charge as a function of angle for bias \SI{300}{\volt} and threshold \SI{1}{\kilo e-}. Plotted only for pixels with multiplication.}
    \label{fig:angular_results}
\end{figure}

In order to evaluate the response of the sensor, a comparison to the analytical prediction values was done -- both for the track length as well as collected charge.
The angular dependency of number of traversed pixels $l_\mathrm{ideal}$ and the deposited charge $Q_\mathrm{ideal}$ can be calculated according to
\begin{multicols}{2}
    \begin{equation}
        l_\mathrm{ideal} = \left\lceil\tan{\left(\alpha\right)}\frac{t}{p}\right\rceil,
    \end{equation}
    \begin{equation}
        Q_\mathrm{ideal} = \Delta E \frac{t}{\cos{\left(\alpha\right)}},
    \end{equation}
\end{multicols}
where $t = \SI{300}{\micro\meter}$ and $p = \SI{55}{\micro\meter}$ represent the sensor thickness and pixel pitch, respectively.
Furthermore $\Delta E \approx \SI{76}{e-\per\micro\meter}$ represents the average electron-hole production of MIPs in \SI{1}{\micro\meter} of silicon.
The resulting plot in \cref{fig:angular_results} indicates a very good agreement between measurement and approximate analytical predictions.
The smaller measured charge compared to the prediction at near-perpendicular angles corresponds to the previously mentioned volcano or screening effects.

\section{Conclusions}
We have shown excellent multiplication uniformity and overall response of the recent iLGAD sensor, bonded to a Timepix3 readout chip.
Our results also indicate a charge screening effect at near-perpendicular angles that is matching previous observation of standard LGAD type devices.

Overall, the gain layer improves the timing performance, further improvement is expected after applying a time-walk correction.
To fully characterise this technology's timing performance, a study using a thinner sensor and a new readout ASIC with better timing precision would have to be made.

\acknowledgments
We would like to thank teams at University of Manchester, University of Glasgow and University of Edinburgh for discussions and provision of the sample.
This work has been partially supported by the STFC grant, UK ST/T002751/1, the testing and evaluation has been done in the scope of the CERN EP R\&D programme and with the use of the CERN SPS beam-test facility.

\bibliographystyle{JHEP}
\bibliography{biblio.bib}

\providecommand{\href}[2]{#2}\begingroup\raggedright\begin{thebibliography}{10}

\bibitem{lgad}
G.~Pellegrini, P.~Fernández-Martínez, M.~Baselga, C.~Fleta, D.~Flores, V.~Greco et~al., \emph{Technology developments and first measurements of low gain avalanche detectors (lgad) for high energy physics applications}, \href{https://doi.org/https://doi.org/10.1016/j.nima.2014.06.008}{\emph{Nuclear Instruments and Methods in Physics Research Section A: Accelerators, Spectrometers, Detectors and Associated Equipment} {\bfseries 765} (2014) 12}.

\bibitem{ionisation}
W.~Maes, K.~{De Meyer} and R.~{Van Overstraeten}, \emph{Impact ionization in silicon: A review and update}, \href{https://doi.org/https://doi.org/10.1016/0038-1101(90)90183-F}{\emph{Solid-State Electronics} {\bfseries 33} (1990) 705}.

\bibitem{timing}
V.~Sola, R.~Arcidiacono, A.~Bellora, N.~Cartiglia, F.~Cenna, R.~Cirio et~al., \emph{Ultra-fast silicon detectors for 4d tracking}, \href{https://doi.org/10.1088/1748-0221/12/02/C02072}{\emph{Journal of Instrumentation} {\bfseries 12} (2017) C02072}.

\bibitem{atlas}
{\scshape {{ATLAS}}} collaboration, \emph{{Technical Proposal: A High-Granularity Timing Detector for the ATLAS Phase-II Upgrade}},  Tech. Rep. \href{https://cds.cern.ch/record/2623663}{CERN-LHCC-2018-023, LHCC-P-012}, CERN, Geneva (2018), \href{https://doi.org/10.17181/CERN.CIUJ.KS4H}{DOI}.

\bibitem{cms}
{\scshape {{CMS}}} collaboration, \emph{{Technical proposal for a MIP timing detector in the CMS experiment Phase 2 upgrade}},  Tech. Rep. \href{https://cds.cern.ch/record/2296612}{CERN-LHCC-2017-027, LHCC-P-009}, CERN, Geneva (2017), \href{https://doi.org/10.17181/CERN.2RSJ.UE8W}{DOI}.

\bibitem{xray}
M.~Andr{\"{a}}, J.~Zhang, A.~Bergamaschi, R.~Barten, C.~Borca, G.~Borghi et~al., \emph{{Development of low-energy X-ray detectors using LGAD sensors}}, \href{https://doi.org/10.1107/S1600577519005393}{\emph{Journal of Synchrotron Radiation} {\bfseries 26} (2019) 1226}.

\bibitem{fill_factor}
N.~Moffat and R.~Bates, \emph{Simulation of the small pixel effect contributing to a low fill factor for pixellated low gain avalanche detectors (lgad)}, \href{https://doi.org/https://doi.org/10.1016/j.nima.2021.165746}{\emph{Nuclear Instruments and Methods in Physics Research Section A: Accelerators, Spectrometers, Detectors and Associated Equipment} {\bfseries 1018} (2021) 165746}.

\bibitem{ilgad}
M.~Carulla, M.~Fernández-García, P.~Fernández-Martínez, D.~Flores, J.~González, S.~Hidalgo et~al., \emph{Technology developments and first measurements on inverse low gain avalanche detector (ilgad) for high energy physics applications}, \href{https://doi.org/10.1088/1748-0221/11/12/C12039}{\emph{Journal of Instrumentation} {\bfseries 11} (2016) C12039}.

\bibitem{timepix3}
T.~Poikela, J.~Plosila, T.~Westerlund, M.~Campbell, M.D.~Gaspari, X.~Llopart et~al., \emph{Timepix3: a 65k channel hybrid pixel readout chip with simultaneous toa/tot and sparse readout}, \href{https://doi.org/10.1088/1748-0221/9/05/C05013}{\emph{Journal of Instrumentation} {\bfseries 9} (2014) C05013}.

\bibitem{SPIDR}
B.~van~der Heijden, J.~Visser, M.~van Beuzekom, H.~Boterenbrood, S.~Kulis, B.~Munneke et~al., \emph{Spidr, a general-purpose readout system for pixel asics}, \href{https://doi.org/10.1088/1748-0221/12/02/C02040}{\emph{Journal of Instrumentation} {\bfseries 12} (2017) C02040}.

\bibitem{telescope}
{\scshape {{CLICdp}}} collaboration, \emph{{Detector technologies for CLIC}}, CERN Yellow Reports: Monographs, {{CDS}} (2019), \href{https://doi.org/10.23731/CYRM-2019-001}{10.23731/CYRM-2019-001}.

\bibitem{corry}
D.~Dannheim, K.~Dort, L.~Huth, D.~Hynds, I.~Kremastiotis, J.~Kröger et~al., \emph{Corryvreckan: a modular 4d track reconstruction and analysis software for test beam data}, \href{https://doi.org/10.1088/1748-0221/16/03/P03008}{\emph{Journal of Instrumentation} {\bfseries 16} (2021) P03008}.

\bibitem{volcano}
M.~Kroupa, S.~Hoang, N.~Stoffle, P.~Soukup, J.~Jakubek and L.S.~Pinsky, \emph{Energy resolution and power consumption of timepix detector for different detector settings and saturation of front-end electronics}, \href{https://doi.org/10.1088/1748-0221/9/05/C05008}{\emph{Journal of Instrumentation} {\bfseries 9} (2014) C05008}.

\bibitem{screening}
E.~Currás, M.~Fernández and M.~Moll, \emph{Gain reduction mechanism observed in low gain avalanche diodes}, \href{https://doi.org/https://doi.org/10.1016/j.nima.2022.166530}{\emph{Nuclear Instruments and Methods in Physics Research Section A: Accelerators, Spectrometers, Detectors and Associated Equipment} {\bfseries 1031} (2022) 166530}.

\end{thebibliography}\endgroup

\end{document}